\def\iscameraready{1}

\documentclass[runningheads]{llncs}
\usepackage[T1]{fontenc}
\usepackage[english=american]{csquotes}
\usepackage{amsmath,amssymb} 
\usepackage{siunitx} 
\usepackage{graphicx}
\usepackage{subcaption}

\usepackage{ragged2e}
\usepackage{rotating}

\usepackage{makecell}
\usepackage{longtable}
\usepackage{array}
\usepackage{multirow}
\usepackage{threeparttable}

\usepackage{calc}

\usepackage[hidelinks]{hyperref}
\usepackage{cleveref}
\usepackage[hyphenbreaks]{breakurl}
\usepackage{xurl}
\hypersetup{breaklinks=true}

\usepackage[acronyms,nonumberlist,nogroupskip,nopostdot]{glossaries}

\usepackage{algorithm2e} 
\SetAlFnt{\small}
\usepackage{listings}

\usepackage[inline]{enumitem}

\usepackage{fontawesome5}

\usepackage[textsize=scriptsize]{todonotes}
\usepackage[
    backend=bibtex,
    urldate=long,
    sorting=none,
    sortcites=true]{biblatex}
    \DefineBibliographyStrings{english}{%
        backrefpage = {cited on page},
        backrefpages = {cited on pages},
}

\newcommand\citeA[1]{\citeauthor*{#1} \cite{#1}}

\usepackage{tikz}
\usepackage[siunitx, RPvoltages]{circuitikz}
\usepackage[linguistics, edges]{forest}

\usetikzlibrary{positioning}
\usetikzlibrary{arrows}
\usetikzlibrary{calc} 
\usetikzlibrary{math}
\usetikzlibrary {shapes.multipart}
\usetikzlibrary {shapes.geometric}
\usetikzlibrary{patterns}
\usetikzlibrary{intersections}

\tikzset{
every text node part/.style={align=center},
every node/.style={outer sep=0pt}, 
text_only/.style={anchor=west},
r_block/.style 2 args={rectangle, rounded corners, text centered, draw=black, fill=white, minimum width = #1, minimum height = #2, anchor=west},
s_block/.style 2 args={rectangle, text centered, draw=black, fill=white, minimum width = #1, minimum height = #2, anchor=west},
memory/.style={rectangle split, rectangle split parts=4, rectangle split part align=center, draw=black, thin, text width=2.5cm, text centered, execute at begin node=\strut, outer sep=2pt}
}

\tikzset{zlevel/.style={%
    execute at begin scope={\pgfonlayer{#1}},
    execute at end scope={\endpgfonlayer}
}}

\tikzset{
    make origin horizontal center of bounding box/.style={%
        execute at end picture={%
            \path
                let 
                    \p1=(current bounding box.west),
                    \p2=(current bounding box.east)
                    in ({-max(-1*\x1,\x2)},\y1) ({max(-1*\x1,\x2)},\y1);
        }
    }
}

\newcolumntype{C}[1]{>{\centering\arraybackslash}m{#1}}
\newcolumntype{V}[1]{>{\centering}m{#1}}
\newcolumntype{P}[1]{>{\RaggedRight\arraybackslash}p{#1}}

\makeatletter
\renewcommand{\fnum@figure}{Figure \thefigure}
\makeatother

\catcode`_=12 %
\renewcommand{\texttt}[1]{%
  \begingroup
  \ttfamily
  \begingroup\lccode`~=`/\lowercase{\endgroup\def~}{/\discretionary{}{}{}}%
  \begingroup\lccode`~=`[\lowercase{\endgroup\def~}{[\discretionary{}{}{}}%
  \begingroup\lccode`~=`.\lowercase{\endgroup\def~}{.\discretionary{}{}{}}%
  \begingroup\lccode`~=`_\lowercase{\endgroup\def~}{_\discretionary{}{}{}}%
  \catcode`/=\active\catcode`[=\active\catcode`.=\active\catcode`_=\active
  \scantokens{#1\noexpand}%
  \endgroup
}
\catcode`_=8 %

\newglossary[slg]{symbols}{sym}{sbl}{List of Symbols} 
\glsaddkey{unit}{\glsentrytext{\glslabel}}{\glsentryunit}{\GLsentryunit}{\glsunit}{\Glsunit}{\GLSunit}\glssetnoexpandfield{unit}
\newglossarystyle{symbunitlong}{%
	\setglossarystyle{super3col}
		{\end{supertabular}}%

}

\newacronym{aes}{AES}{advanced encryption standard}
\newacronym{isa}{ISA}{instruction set architecture}
\newacronym[longplural={cache timing vulnerabilities}]{ctv}{CTV}{cache timing vulnerability}
\newacronym{llc}{LLC}{last-level cache}
\newacronym{lru}{LRU}{least recently used}
\newacronym{plru}{PLRU}{pseudo-LRU}
\newacronym{fifo}{FIFO}{first-in first-out}
\newacronym{lfu}{LFU}{least frequently used}
\newacronym{msb}{MSB}{most significant bit}
\newacronym{lsb}{LSB}{least significant bit}
\newacronym{des}{DES}{data encryption standard}
\newacronym[shortplural={S-boxes},longplural={substitution boxes}]{sbox}{S-box}{substitution box}
\newacronym{tlb}{TLB}{translation look-aside buffer}
\newacronym{cpu}{CPU}{central processing unit}
\newacronym{sbc}{SBC}{single-board computer}
\newacronym{risc}{RISC}{reduced instruction set computer}
\newacronym{tee}{TEE}{Trusted Execution Environment}
\newacronym{smt}{SMT}{simultaneous multithreading}
\newacronym{ctvs}{CTVS}{cache timing vulnerability score}

\makeglossaries

\begin{document}

\newcommand\authNameZero{Cédrick Austa}
\newcommand\authOrcidZero{0009-0002-1223-3285}
\newcommand\authEmailPrefixZero{cedrick.austa}
\newcommand\authEmailSuffixZero{@ulb.be}
\newcommand\authInstitutionZero{Université Libre de Bruxelles, Av. Roosevelt 50, 1050 Bruxelles, Belgium}
\newcommand\authNameOne{Jan Tobias Mühlberg}
\newcommand\authOrcidOne{0000-0001-5035-0576}
\newcommand\authEmailPrefixOne{jan.tobias.muehlberg}
\newcommand\authEmailSuffixOne{@ulb.be}
\newcommand\authInstitutionOne{École Polytechnique de Bruxelles, Université Libre de Bruxelles, Av. Roosevelt 50, 1050 Bruxelles, Belgium}
\newcommand\authNameLast{Jean-Michel Dricot}
\newcommand\authOrcidLast{0000-0002-8539-9940}
\newcommand\authEmailPrefixLast{jean-michel.dricot}
\newcommand\authEmailSuffixLast{@ulb.be}
\newcommand\authInstitutionLast{École Polytechnique de Bruxelles, Université Libre de Bruxelles, Av. Roosevelt 50, 1050 Bruxelles, Belgium}

\title{Systematic Assessment of Cache Timing Vulnerabilities on RISC-V Processors}
\titlerunning{Cache Timing Vulnerabilities on RISC-V Processors}

\author{
    \authNameZero\inst{1}\orcidID{\authOrcidZero} \and \authNameOne\inst{1}\orcidID{\authOrcidOne} \and \authNameLast\inst{1}\orcidID{\authOrcidLast}
}
\authorrunning{C. Austa et al.}
\institute{
    \authInstitutionZero\\
    \email{\{\authEmailPrefixZero,\authEmailPrefixOne,\authEmailPrefixLast\}\authEmailSuffixZero}
}
\maketitle

\begin{abstract}
While interest in the open
RISC-V instruction set architecture is growing, tools to assess the
security of concrete processor implementations are lacking.  There are
dedicated tools and benchmarks for common microarchitectural side-channel
vulnerabilities for popular processor families such as Intel x86-64 or ARM,
but not for RISC-V. In this paper we describe our efforts in porting an
Intel x86-64 benchmark suite for cache-based timing vulnerabilities to RISC-V.
We then use this benchmark to evaluate the security of three commercially
available RISC-V processors, the T-Head C910 and the SiFive U54 and U74
cores. We observe that the C910 processor exhibits more distinct timing
types than the other processors, leading to the assumption that code
running on the C910 would be exposed to more microarchitectural
vulnerability sources.
In addition, our evaluation reveals that $65.9\%$ of the vulnerabilities
covered by the benchmark exist in all processors, while only $6.8\%$ are
absent from all cores. Our work, in particular the ported benchmark, aims
to support RISC-V processor designers to identify leakage sources early in
their designs and to support the development of countermeasures.


\end{abstract}


\section{Introduction}

Modern processors include many performance-enhancing features, such as
cach\-ing, paging, out-of-order execution and speculative executions, which improve
computer system performances.  However, these features can be exploited to
create side channels.

A \textit{side channel} is an unintended communication channel between two
entities, where one party observes information that is inadvertently leaked
by the functioning of the other party's system~\cite{Szefer2019}. This
communication channel can be exploited either actively, by interacting with
the system and observing how it reacts, or passively, by observing
behavioral changes.
A processor microarchitecture is a processor-specific logic implementation
of an \gls{isa}~\cite{Harris2022}.
While the latter is the definition of the instruction set requirements,
e.g., the supported instruction set and data types, the registers, and the
privilege modes,
the microarchitecture describes the specifics of an \gls{isa}
implementation,
e.g.,
the instruction pipeline stages,
the implementation of the cache and other optimization features.
Microarchitectural side-channel vulnerabilities stem from and exploit
effects of operations performed on a specific processor
microarchitecture, which are not documented in the \gls{isa}.
Examples of such vulnerabilities are cache-based side-channels
which are used to extract information by observing operations
on the cache or to build more sophisticated attacks, e.g., Meltdown
vulnerability~\cite{Lipp2018}.  This category of
vulnerabilities has been extensively studied for Intel
processors~\cite{Deng2020a,Rauscher2025,Purnal2023a}, AMD
processors~\cite{Ren2021,Irazoqui2016,Lipp2020,Purnal2023a}, and
ARM processors~\cite{Lipp2016,Deng2021a}, and tools have been made
available to evaluate their security.

The recent and open-source RISC-V \gls{isa}, which has gained a lot of
popularity amongst industry and researchers, did not receive that much
attention.
Though, we know that RISC-V implementations are not exempt from
microarchitectural vulnerabilities~\cite{Gerlach2023,Le2023}.  E.g.,
\citeA{Gerlach2023} already demonstrated that some differences may be
observed between vulnerabilities on the Intel x86-64 architecture and the
RISC-V architecture.  Moreover, since RISC-V is an open-source \gls{isa},
different RISC-V hardware implementations are made available by CPU
vendors, which leads to processor-specific vulnerabilities.  As a result,
the RISC-V architecture needs its own security toolkit to evaluate existing
and future RISC-V processors. Recently, \citeA{Thomas2024} provides
automatic detection of architectural vulnerabilities but benchmarks or test
suites to assess the microarchitectural attack surface of a given RISC-V
implementation are currently missing. In this paper we address this
shortcoming. With a focus on cache-based timing vulnerabilities, we ported
a benchmark~\cite{Deng2020a,Deng2019} to target RISC-V and evaluate it on
three commercially available processors.

\paragraph{Contributions.}
\label{sec:intro-contrib}

We ported and extended a benchmark from Intel x86-64~\cite{Deng2020a} to detect
cache-based side-channel vulnerabilities for the RISC-V \gls{isa}. The original
benchmark is based on a theoretical three-step model proposed
in~\citeA{Deng2019} and is made of two software components: the first allows to
observe the different cache timing types on a processor, while the second
allows to determine cache-based timing vulnerabilities on the L1 data cache of
this processor.  By allowing for different cache hierarchies, different cache
set associativities, and different cache eviction strategy parameters, our ported
benchmark is more flexible and easier to adapt to new RISC-V implementations.
Our contributions are:
\begin{itemize}
    \item porting and translating original work from~\citeA{Deng2020a} to
        RISC-V processors;
    \item refactoring the benchmark to support new RISC-V processor
        configurations;
    \item evaluating cache-based vulnerabilities on the three commercially
        available RISC-V processors, the T-Head C910, and the SiFive U74 and
        U54;
    \item identifying and interpreting timing differences between memory
        operations;
    \item identifying L1 data cache-based vulnerabilities across all
        implementations.
\end{itemize}
With our evaluation, we show that the translated benchmark can be used to
successfully identify cache timing leakage sources and potential
vulnerabilities on RISC-V processors.  By doing so, we demonstrate the utility
of such a benchmark for RISC-V chip designers to identify vulnerabilities in
their designs  for which countermeasures should be implemented.  Our results
show that 39 out of the 88 cache-based timing vulnerabilities highlighted
in~\citeA{Deng2020a} are present in all implementations. We make our benchmark
available under an open-source license:
\url{https://github.com/ReSP-Lab/risc-v-sidechannel-benchmark}

Even though such an evaluation was never performed on RISC-V
implementations, we do not consider that following confidential disclosure
procedures with the manufacturers are warranted:
\begin{enumerate*}[label=(\roman*)]
  \item{most of the implementations we evaluate are used in development
boards that are mainly dedicated to hobbyists and researchers,}
  \item{the benchmark on which we based our work as well as the covered
vulnerabilities are not new,}
  \item{the vulnerabilities are rather involved and  not easily exploitable
in practice, and}
  \item{subsets of these vulnerabilities have already been reported
by~\citeA{Gerlach2023} for the same or similar processors.}
\end{enumerate*}

\section{Background}\label{sec:bg}

Below we provide background on caches and caching behavior in modern
processors, we outline how caches can be abused in side-channel attacks,
and we summarize \citeA{Deng2020a}'s work as the foundation of our
research.

\subsection{System Caches and Caching}
\label{sec:bg:defs}

To optimize performance, when a processor requires data from (slow) main
memory it first checks the (fast) caches if this data is already available.
If the data is in the cache, the cache \textit{hits}; otherwise, the cache
\textit{misses} and data is fetched from main memory and placed into the
cache.  \textit{Caching} exploits \textit{temporal locality} and
\textit{spatial locality}~\cite{Harris2022}: As a consequence of temporal
locality, data is loaded into the cache from main memory when a cache miss
occurs. This happens in \textit{cache blocks} that include data from
adjacent addresses, thus exploiting spatial locality.

\paragraph{Cache Hierarchy.}
To decrease the \textit{miss penalty}, i.e., the latency overhead due to
cache misses, a \textit{multi-level cache hierarchy} is often used.  The
highest level and the lowest level in the hierarchy respectively are the L1
cache and the \textit{\gls{llc} cache}.  The lower the cache level in the
hierarchy, the larger are both the cache size and the latency to
access data in this cache.

\paragraph{Cache Specifications.}
The performance impact of a cache depends on cache capacity $C$, block size
$b$, and the degree of associativity $N$~\cite{Harris2022}.
Let $B=C/b$ be the number of cache blocks in a cache.
The cache is \textit{$N$-way set associative} when these blocks are grouped
into $S$ sets of $N$ ways, i.e., block locations in the set,
with $S=B/N$ and $1 \le N \le B$.
The number of ways per set determines how memory addresses in main memory are
mapped to cache locations:
If $N=1$, any memory address is mapped to only one block; such a cache is
called a \textit{direct mapped} cache.
If $N=B$, any memory address can be mapped to every blocks in the cache; such a
cache is called a \textit{fully associative} cache.

\paragraph{Cache Replacement Policy.}
When a miss occurs while the cache is fully populated, a cache block needs to be evicted to load the
target block into the cache set.
For a $N$-way set associative cache with $N > 1$, a
\textit{cache replacement policy} is necessary to select which block to evict.
Examples of cache replacement policies are the
\textit{random} replacement policy (evicting a random block), the
\gls{fifo} replacement policy (evicting the oldest
block), or the \gls{lru} policy.
In \gls{lru}, to avoid tracking the last use of each way in a set,
approximations of this policy,
so-called \gls{plru} policies, are used.

\paragraph{Cache Coherence Protocol.}
In multicore processors, cache coherence protocols are used
to track cache block states among private and shared caches.
By doing so, each core reads the same content from a cache block at any time
between two write operations.
The two main categories of cache coherence protocols are
\textit{directory-based} protocols and \textit{snooping} protocols.
The former track cache block states using a directory, in a centralized
way where
the state of a cache block is known by accessing the directory.
The latter track cache block states using a snooping bus and
the state of a cache block is stored locally in the cache and each cache
block state alteration is broadcasted on the bus for other listeners.
Examples of states are the followings: M(odified), O(wned), E(xclusive),
S(hared), or I(nvalid).
The choice of a cache coherence protocol has impact on performance
and scalability.

\subsection{Cache-based Timing Vulnerabilities}\label{sec:bg:ctvs}

Cache-based microarchitectural attacks exploit the cache status as a
leaking information source.  Information is collected either by observing
timing differences on cache block accesses~\cite{Bernstein2005,Lipp2018} or
by measuring power consumption differences on cache block content
change~\cite{Bertoni2005,Kogler2023}.
\textit{Cache-based timing} attacks exploit the observed latency between
cache hits and cache misses, e.g. the access time increases whenever cache
blocks to load are lower in the cache hierarchy.  Attackers can interact
with the cache using a specific memory operation and measure the operation
time or latency to infer the victim's cache state.
According to Su et al.~\cite{Su2021a}, a succeeding cache attack depends on the
following three conditions:
\begin{enumerate}
    \item a relation should exist between a change in the cache state and
        target sensitive information;
    \item at least one cache in the cache hierarchy is shared between both
        the attacker program and the victim program;
    \item both programs sharing the cache can infer changes to each other's
        cache state by monitoring their own cache state.
\end{enumerate}
To illustrate how the cache state may be exploited,
we describe selected well-known cache-based timing attacks below.

\paragraph{Prime$+$Probe~\cite{Osvik2006}.}
This first cache-based timing attack works on cache sets as follows:
\begin{enumerate*}[label=(\roman*)]
    \item the attacker fills one or more sets with its own cache blocks
        (\textit{prime}),
    \item victim cache blocks are accessed, and
    \item the attacker reloads its own cache blocks in each set and measures
        timing (\textit{probe}). 
\end{enumerate*}
In this attack, the cache probing is longer if attacker cache blocks were
evicted by victim cache blocks.

\paragraph{Evict$+$Time~\cite{Osvik2006}.}
This attack, which was introduced with Prime$+$Probe, consists in the following
steps:
\begin{enumerate*}[label=(\roman*)]
    \item the attacker runs the victim program and measures timing,
    \item the attacker evicts a victim cache block (\textit{evict}), and
    \item the attacker runs the victim program again and measures timing
        (\textit{time}).
\end{enumerate*}
If the evicted cache block was accessed by the victim, the execution time is
longer to load the victim cache block.

\paragraph{Flush$+$Reload~\cite{Yarom2014}.}
To perform this attack,
\begin{enumerate*}[label=(\roman*)]
    \item the attacker flushes a victim cache block (\textit{flush}),
    \item the victim program eventually accesses the victim cache block, and
    \item the attacker reloads the victim cache block and measures timing
        (\textit{reload}).
\end{enumerate*}
If the victim accessed the flushed cache block, the reload operation time is
shorter for the attacker.
A variant of this attack, \textit{Evict$+$Reload}, was proposed by
\citeA{Gruss2015} to avoid the flush instruction requirement.

\paragraph{Flush$+$Flush~\cite{Gruss2016}.} This attack uses steps
\begin{enumerate*}[label=(\roman*)]
    \item and \item from the Flush$+$Reload attack.
    For step \item the attacker flushes the victim cache block again and
        measures timing.
\end{enumerate*}
If the victim accessed their cache block, the flush operation takes longer
than if they did not.


\begin{table*}[ht!]
\setlength\extrarowheight{4pt}
\caption{\small
L1 cache timing-based vulnerabilities~\cite{Deng2020a}.
The {\em No.} column assigns each type of vulnerability a number range.
The {\em Attack Strategy} column gives a common name for each set of
vulnerabilities and each vulnerability types
that would be exploited in an attack in a similar manner.
Inv. means invalidation.
}
\label{table:vulnerabilities}
\begin{minipage}{0.50\textwidth}
\begin{subtable}[t]{\textwidth}
\centering
{
\resizebox{1\textwidth}{!}{%
\scriptsize
    \begin{tabular}[t]{|C{1cm}|C{4.5cm}|}
    \hline

    \shortstack{No.}
    & \shortstack{Attack Strategy}\\ \hline\hline

    1-4 &\shortstack{Cache Collision} \\ \hline
    5-8 & \shortstack{Flush + Reload}\\ \hline
    9-10 & \shortstack{Reload + Time}\\ \hline
    11-14 & \shortstack{Flush + Probe}\\ \hline
    15-16 & \shortstack{Flush + Time}\\ \hline
    17-20 & \shortstack{Cache Coherence\\ Flush + Reload}\\ \hline
    21-28 & \shortstack{Cache Coherence\\ Prime + Probe}\\ \hline
    29-32 & \shortstack{Cache Coherence\\ Evict + Time}\\ \hline
    33-36 & \shortstack{Bernstein's Attack}\\ \hline
    37-38 & \shortstack{Evict + Probe}\\ \hline
    39-40 & \shortstack{Prime + Time}\\ \hline
    41-42 & \shortstack{Evict + Time}\\ \hline
    43-44 & \shortstack{Prime + Probe}\\ \hline

    \end{tabular}
    } 
} 
\caption{\small 
Third step as memory access operation.}
\label{table:table_vuln_a}
\end{subtable}
\end{minipage}
\begin{minipage}{0.50\textwidth}
\begin{subtable}[t]{\textwidth}
\centering
{
\resizebox{1\textwidth}{!}{%
\scriptsize
\begin{tabular}[t]{|C{1cm}|C{4.5cm}|}
    \hline

    \shortstack{No.}
    & \shortstack{Attack Strategy}\\ \hline \hline

    45-46 & \shortstack{Cache Collision Inv.} \\ \hline
    47-50 & \shortstack{Flush + Flush} \\ \hline
    51-52 & \shortstack{Flush + Reload  Inv.} \\ \hline
    53-54 & \shortstack{Reload + Time  Inv.} \\ \hline
    55-58 & \shortstack{Flush + Probe Inv.} \\ \hline
    59-60 & \shortstack{Flush + Time  Inv.} \\ \hline
    61-64 & \shortstack{Cache Coherence\\  Flush + Reload Inv.} \\ \hline
    73-76 & \shortstack{Cache Coherence\\ Evict + Time Inv.} \\ \hline
    77-80 & \shortstack{Bernstein's Inv. Attack} \\ \hline
    81-82 & \shortstack{Evict + Probe Inv.} \\ \hline
    83-84 & \shortstack{Prime + Time Inv.} \\ \hline
    85-86 & \shortstack{Evict + Time Inv.} \\ \hline
    87-88 & \shortstack{Prime + Probe Inv.} \\ \hline

    \end{tabular}
    } 
} 
\caption{\small
Third step as invalidation operation.}
\label{table:table_vuln_b}
\end{subtable}
\end{minipage}
\end{table*}

\begin{table}[ht!]
    \caption{\small Test configurations for the L1-D cache benchmarks.
        For memory accesses: \texttt{read} and \texttt{write} operations.
        For invalidations: \texttt{flush} and \texttt{write} operations.}
    \label{table:test_config}
    \begin{center}
        \small
        \resizebox{0.95\textwidth}{!}{%
        \begin{tabular}[c]{
                |C{0.2\linewidth}
                ||C{0.2\linewidth}
                |C{0.2\linewidth}
                |C{0.2\linewidth}
                |C{0.2\linewidth}|}
            \hline
            \multicolumn{1}{|C{0.2\linewidth}||}{\textbf{Test config.}} &
            \multicolumn{1}{C{0.2\linewidth}|}{\textbf{Step 1}} &
            \multicolumn{1}{C{0.2\linewidth}|}{\textbf{Step 2}} &
            \multicolumn{1}{C{0.2\linewidth}|}{\textbf{Step 3}} &
            \multicolumn{1}{C{0.2\linewidth}|}{\textbf{Run}}
            \\
            \hline
            \hline
            \texttt{RF_RF_RF_TS} & read/flush & read/flush & read/flush
                & time-slicing \\
            \hline
            \texttt{RF_RF_RF_SMT} & read/flush & read/flush & read/flush
                & \acrshort{smt} \\
            \hline
            \texttt{RF_RF_W_TS} & read/flush & read/flush & write & time-slicing \\
            \hline
            \texttt{RF_RF_W_SMT} & read/flush & read/flush & write & \acrshort{smt} \\
            \hline
            \texttt{RF_W_RF_TS} & read/flush & write & read/flush & time-slicing \\
            \hline
            \texttt{RF_W_RF_SMT} & read/flush & write & read/flush & \acrshort{smt} \\
            \hline
            \texttt{RF_W_W_TS} & read/flush & write & write & time-slicing \\
            \hline
            \texttt{RF_W_W_SMT} & read/flush & write & write & \acrshort{smt} \\
            \hline
            \texttt{W_RF_RF_TS} & write & read/flush & read/flush & time-slicing \\
            \hline
            \texttt{W_RF_RF_SMT} & write & read/flush & read/flush & \acrshort{smt} \\
            \hline
            \texttt{W_RF_W_TS} & write & read/flush & write & time-slicing \\
            \hline
            \texttt{W_RF_W_SMT} & write & read/flush & write & \acrshort{smt} \\
            \hline
            \texttt{W_W_RF_TS} & write & write & read/flush & time-slicing \\
            \hline
            \texttt{W_W_RF_SMT} & write & write & read/flush & \acrshort{smt} \\
            \hline
            \texttt{W_W_W_TS} & write & write & write & time-slicing \\
            \hline
            \texttt{W_W_W_SMT} & write & write & write & \acrshort{smt} \\
            \hline
        \end{tabular}
        }
    \end{center}
\end{table}

\subsection{\citeA{Deng2020a} Benchmark Suite}
\label{sec:bg:deng}

\subsubsection{Observable timing types.}
In their work, \citeA{Deng2020a} study timing differences between memory
operations depending on the cache level in which data is located, depending
on the cache block data state (i.e., \textit{dirty} or \textit{clean})
and whether the attacker is running their program on the same core
as the victim.
As a result, 66 different timing types are considered instead of only
the two traditional timing types for cache-based timing vulnerabilities
(i.e., \textit{fast} or \textit{slow}).
Even though some of these timing types were already exploited in the
literature, e.g., both Flush$+$Reload~\cite{Yarom2014} and
Flush$+$Flush~\cite{Gruss2016} rely on at least two of these 66 timing types,
their work emphasizes the existence of such distinct timing types.
To obtain the 66 timing types (i.e., 22 timing types per operation)
the following cases are considered:
\begin{itemize}
    \item either a read, write or flush operation is used to access
        data (3 operations);
    \item data is present either on the local L1/L2/L3 cache or
        on a remote L1/L2/L3 cache and the cache block in which
        data is present is either in dirty or clean state
        (6 $\times$ 2 $=$ 12 cases),
        \textbf{or}
        data is present both in either the local L1, L2 or L3
        cache and either the remote L1, L2 or L3 cache in clean state
        (3 $\times$ 3 $=$ 9 cases),
        \textbf{or}
        data is only present in the DRAM (1 case).
\end{itemize}
A first part of their benchmark suite implementation generates histograms to
identify the different timing types existing on a target.
Each timing type which may be distinguished from the others in this histogram
corresponds to a timing leakage source which can be exploited by an adversary
to monitor cache usage.

\subsubsection{Three-step model.}
\citeA{Deng2020a} propose an improvement for a theoretical model~\cite{
    Deng2019}
describing cache timing vulnerabilities as sequences of three steps
(i.e., three memory-related operations), each modifying the state of a target
cache block:
state initialization, state alteration, and timing observation.
From each step, one of 17 possible cache block states should be reached
for the target cache block.
The memory-related operations which can be used for each step correspond
to the \texttt{read}, \texttt{write} and \texttt{flush} operations, depending
on the resulting cache block state.

To validate their three-step model and detect all possible
L1-D cache-based timing vulnerabilities on concrete processors,
a subset of the previous timing types are used in a benchmark suite.
In particular, benchmarks are automatically generated to evaluate the
88 vulnerabilities \citeA{Deng2020a} identified as ``strong''
out of 4913 possible vulnerabilities.
These 88 vulnerabilities are summarized in \Cref{table:vulnerabilities}.
For more details,
the reader is re-oriented towards \citeA{Deng2020a}.

Moreover, to evaluate the existence of each vulnerability on a processor,
their benchmark suite considers 16 test configurations for each
vulnerability.
These configurations aim to reach the same three cache block states for a
vulnerability by applying different memory-related operation;
cache block access can be performed with \texttt{read} or \texttt{write}
operations, while cache block invalidation can be performed with
\texttt{flush} or \texttt{write}.
These configurations also compare timing differences observed
whenever a victim and an adversary are located on the same physical core
and whenever either \textit{time-slicing} or \textit{\gls{smt}} is used.
With some of these configurations, \citeA{Deng2020a} identified vulnerabilities
relying on cache coherence.
\Cref{table:test_config} lists these test configurations.

\section{Threat Model, Benchmark Suite, Experimental Setup}
\label{sec:methodology}

The goal of our work is to develop and evaluate a benchmark to assess the
security of RISC-V implementations regarding cache-based timing
vulnerabilities. Below we define our attacker model, and present the ported
benchmark suite and our experimental setup.

\subsection{Threat Model}

We assume an attacker model where the adversary is able to gain
unprivileged access to a system of interest and execute code on that
system, and that this code can access hardware performance counters of the
system, in particular clock cycle counters to measure memory operation
latencies.
Moreover, we only consider vulnerabilities which are evaluated in the
benchmark (cf.~\Cref{sec:bg:deng}, \Cref{table:vulnerabilities}).
We also make the
assumptions that the conditions listed in \Cref{sec:bg:ctvs} are satisfied:
the existence of a link between cache states and victim secrets, and a
shared and mutually observable cache state between victim and attacker
code.

Under these conditions, we can rely on measurements from hardware
performance counters to determine if the system is vulnerable or not.  If
no vulnerability is found that way---i.e., by relying on accurate clock cycle
counters---the chances for an adversary to find vulnerabilities by
relying on less accurate timers, such as POSIX timers, would be null.
However, vulnerabilities found that way may not always result in
exploitable vulnerabilities in practice. Verifying exploitability would
typically be specific to a victim program and goes beyond the scope of this
work. Also the inclusion of additional sources of information leakage,
hardware side channels or microarchitectural side channels other than
caches, is beyond the scope of this work.

\subsection{Benchmark Suite Implementation \& Porting}
\label{sec:porting}

Regarding the porting to RISC-V, we decided
\begin{enumerate*}[label=(\roman*)]
    \item to segment the source code into smaller dedicated header and source
        files to ease for the code maintenance;
    \item to move all inline assembly snippets into inline assembly functions
        in a dedicated file to reduce the number of redundant code lines and,
        hence, to ease source code maintenance and corrections;
    \item to dedicate a header file for each target which defines
        all target specifications, related parameters (e.g., for
        \citeA{Gruss2016b} eviction algorithm), and the (optional) related
        inline assembly \texttt{flush} function, to ease the support
        implementation for a new target;
    \item to allow the user to define which cache levels are exist on the
        target so that we can use the benchmark more constrained or developed
        targets.
\end{enumerate*}

In order to facilitate the evaluation of cache-based timing vulnerabilities,
huge pages of 2MB are used to find conflicting addresses.
Since most processors use virtually indexed physically tagged caches,
this allows us to avoid taking virtual address translation into account.
While this approach is relevant for embedded systems,
it is less realistic for scenarios involving a fully-fledged operating
system.
Yet, since the goal of this work is not to evaluate the difficulty to build
eviction sets on the targets, we consider this a minor limitation.
In this context, eviction is not performed by building conflict sets,
as it is the case in the original benchmark suite~\cite{Deng2020a},
but by using the eviction algorithm presented by \citeA{Gruss2016b}.
\citeA{Deng2020a} only covered processors that use an
\gls{lru} policy on L1 and L2 caches, which means that only sequential
access to $N$ cache blocks is required to evict a cache set.
The authors built conflict sets only to evict the L3 cache of their targets.
When considering targets that use different cache replacement
policies per cache level, this method would require us to build a specific
conflict set for each of these cache levels.
Instead, the eviction algorithm proposed by~\citeA{Gruss2016b} allows us to deal with
different replacement policies than the \gls{lru} policy by adapting the
algorithm parameters, and conflicting addresses are easy to list since the set
index is straightforward to obtain using huge pages on our targets, which
has performance benefits.
Nevertheless, this approach requires to define adequate algorithm
parameters. For our study we manually explored different sets of parameters
and settled on parameters that produce stable results with small error
margins in our evaluation.

\subsection{Evaluation Targets \& Experimental Setup}

For the experiments, three \glspl{sbc} with RISC-V application cores
were used.
These \glspl{sbc} are the BeagleV-Ahead~\cite{BVAhead} (SoC: TH1520, core: C910),
the BeagleV-Fire~\cite{BVFire} (SoC: U54-MC, app. core: U54) and the
HiFive Unmatched~\cite{Unmatched} (SoC: FU740, app. core: U74).
Their cache specifications are given in \Cref{tab:sbc-cache}.
Note that since the cache replacement policy for the L2 cache from U54 is
not documented, we made the assumption it was a \gls{plru} policy.
Experiments were performed with the Ubuntu images available for these
\glspl{sbc}
(images from the \gls{sbc} provider or from Ubuntu Boards).

\begin{table}[t!]
    \caption{Evaluated system-on-chip cache specifications.
        Both SiFive cores use a directory-based cache coherence protocol,
        while the C910 core uses a snooping cache coherence protocol.}
    \label{tab:sbc-cache}
    \begin{center}
        \begin{tabular}[c]{|c|ccc|ccc|}
            \hline
            \multirow{2}{*} {\textbf{Core}}
            & \multicolumn{3}{c|} {\textbf{L1-D cache}}
            & \multicolumn{3}{c|} {\textbf{L2 cache}}\\
            & \textbf{Size} & \textbf{Ways} & \textbf{Replacement}
            & \textbf{Size} & \textbf{Ways} & \textbf{Replacement} \\
            \hline
            \textbf{C910} & 64kB & 2 & FIFO & 1MB & 16 & FIFO \\
            \textbf{U54} & 32kB & 8 & random & 2MB & 16 & PLRU \\
            \textbf{U74} & 32kB & 8 & random & 2MB & 16 & random \\
            \hline
        \end{tabular}
    \end{center}
\end{table}

First, none of the selected \glspl{sbc} benefits from an L3 cache.
Thus, 27 timing types are not considered during the evaluation of memory
operations, resulting to the measurement of only 39 out of the 66 timing types
proposed by~\citeA{Deng2020a}.
Moreover, previously selected images allow the user-mode to use the
\texttt{rdcycle} instruction which are used to obtain high accuracy timings
for the benchmarks.
Unfortunately, by using these images, flush instructions are not available
except for the C910 processor which leaves it available to user-mode programs.
As a result, only 26 timing types can be measured for the evaluations of the
U54 and U74 processors.
However, to be able to measure DRAM latency for each operation,
flush instruction is replaced by L2 eviction when necessary.

Finally, as proposed by~\citeA{Deng2020a}, we measure latencies on 8 cache
lines from distinct cache sets at a time, in order to minimize false negatives.
Regarding false positives, \citeA{Deng2020a} proposed to isolate cores to
reduce software noise. We decided to not isolate cores to have a more realistic
situation, since our configuration is not running anything else than the
operating system and our benchmarks. Moreover, we only observe a small
differences when isolating cores; we
still observe differences between distinct evaluations with isolation, probably
due to last level cache conflicts with the operating system and our process,
and due to random cache policies.

\section{Results}
\label{sec:discussion}

\subsection{Timing Types}
\label{sec:disc:timing}

\subsubsection{Measurements.}
\begin{figure}[ht]
    {\centering
        \begin{subfigure}{0.49\textwidth}
            \centering
            \includegraphics[width=\textwidth]{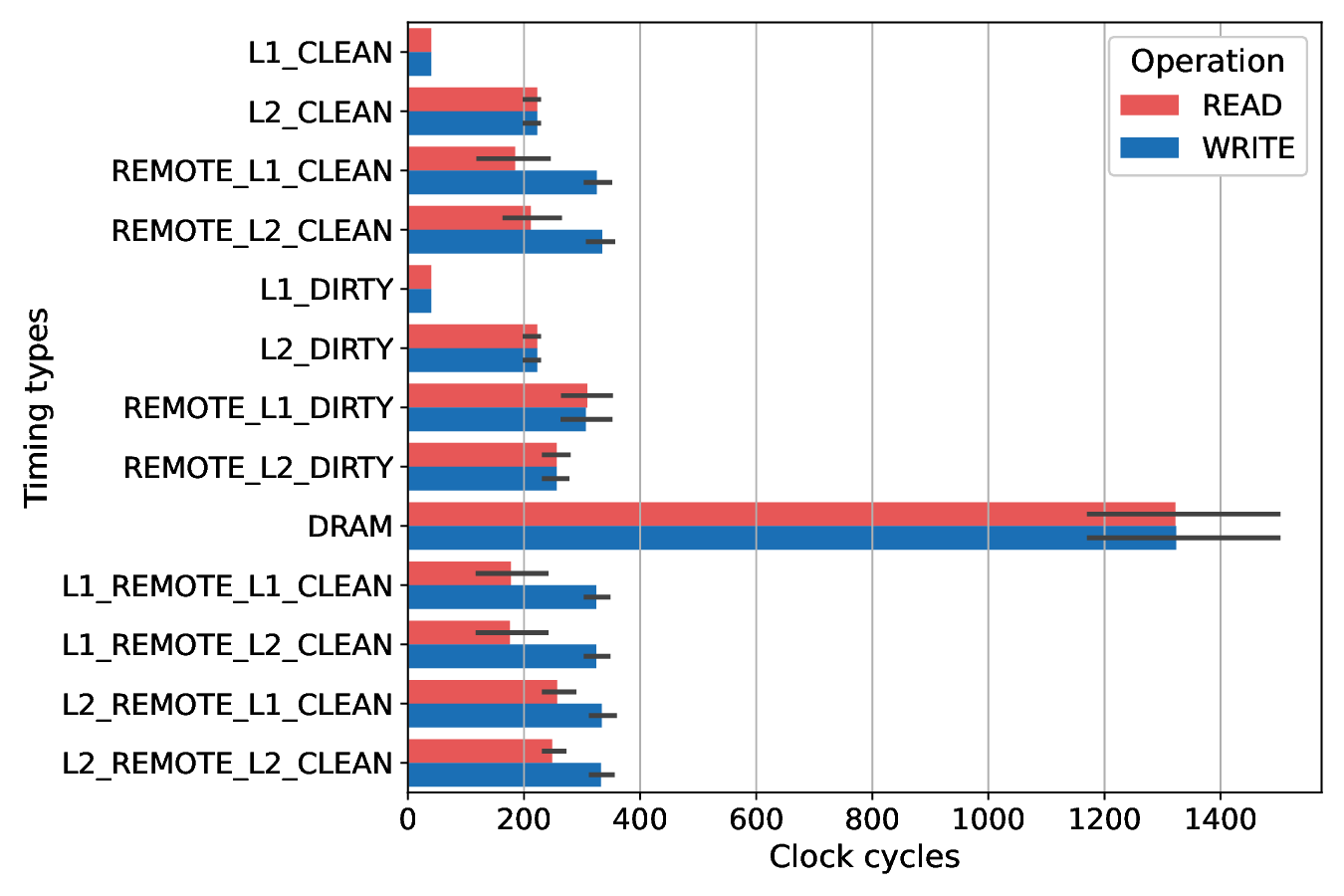}
            \caption{U74}
        \end{subfigure}
        \begin{subfigure}{0.49\textwidth}
            \centering
            \includegraphics[width=\textwidth]{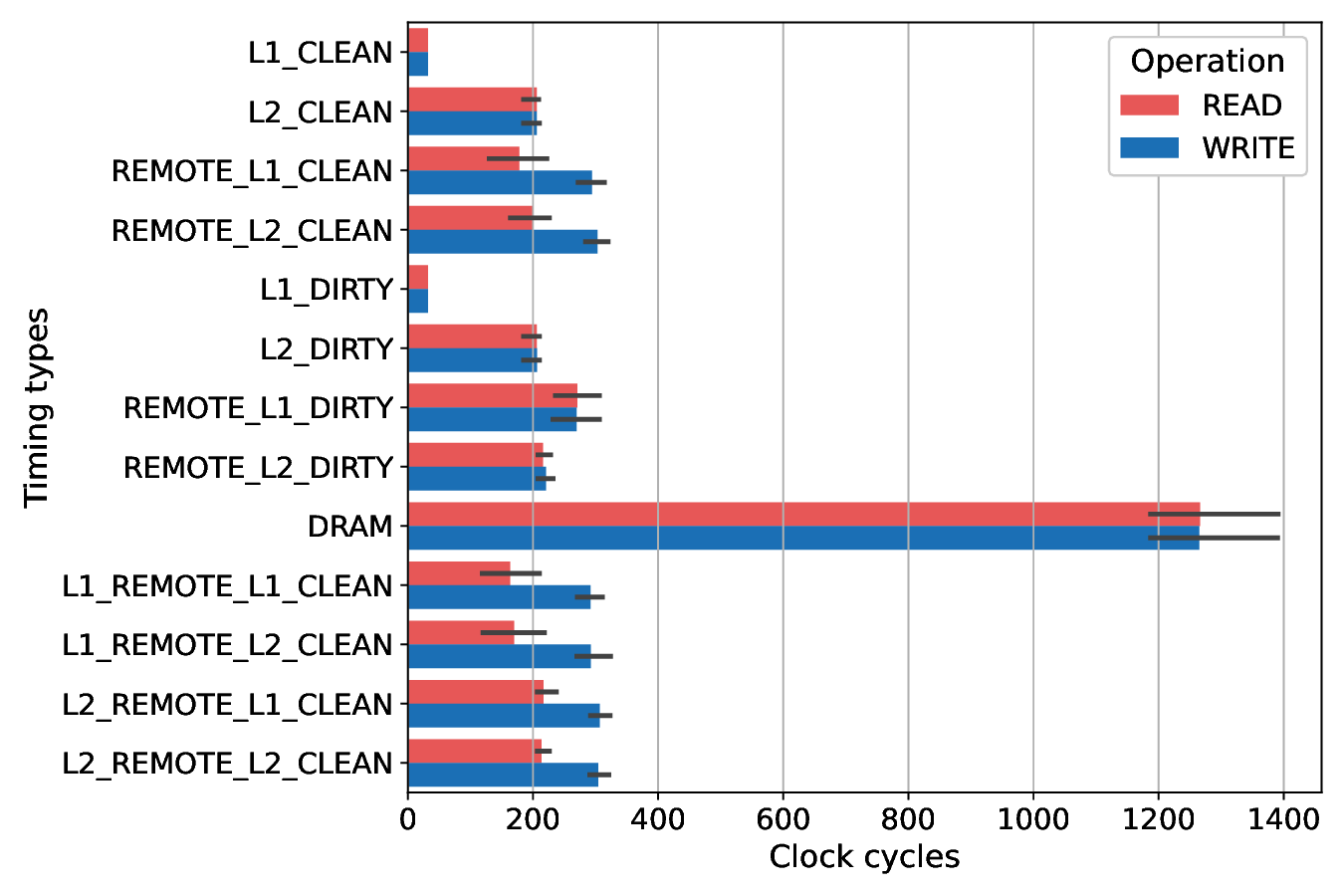}
            \caption{U54}
        \end{subfigure}
        \begin{subfigure}{0.52\textwidth}
            \centering
            \includegraphics[width=\textwidth]{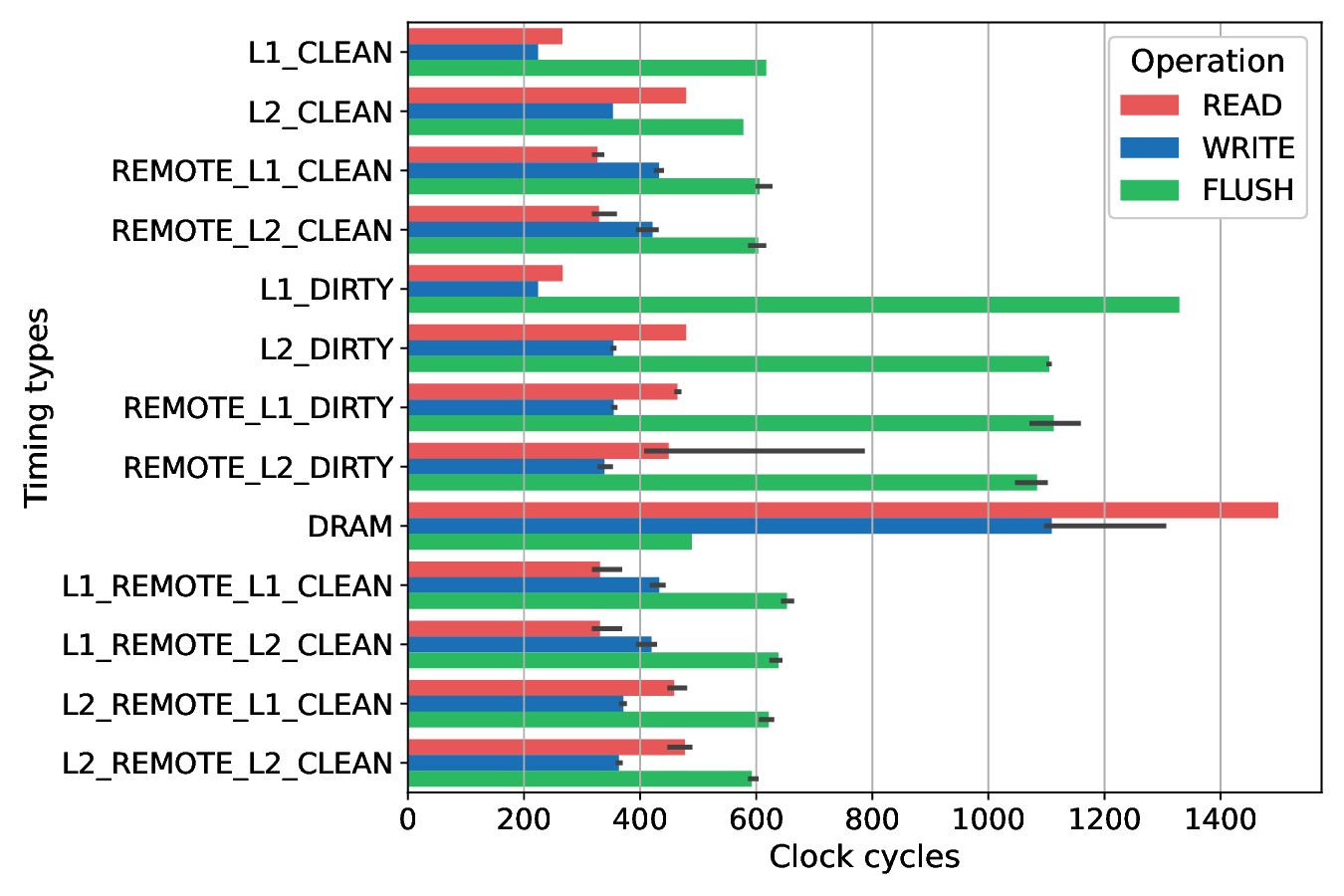}
            \caption{C910}
        \end{subfigure}
    \caption{Most frequent clock cycle latencies over 10000 tests, per timing
        type and per target.
        Label prefixes \texttt{Lx_} or \texttt{REMOTE_Lx}, refer
        to the cache block location in the cache level \texttt{x} on local or
        remote core.
        Label suffixes \texttt{_CLEAN} or \texttt{_DIRTY}, refer
        to the cache block state before the memory operation, respectively.
    }
    \label{fig:histograms}
    } 
\end{figure}

The observed timing types from our tests on the three evaluated cores are
given in \Cref{fig:histograms}.
The most frequent clock cycle latency from all tests
is presented as a bar plot
and a percentile interval of 95\% is superposed to the plot as an error bar.
Timing type labels refer to a subset of the 22 timing types per operation
described in \Cref{sec:bg:deng}.
Obviously, timing types related to L3 cache are not considered.

\subsubsection{Observations and Preliminary Discussion.}
From \Cref{fig:histograms}, the observed clock cycle latencies
differ between most of the evaluated timing types, for all cores.
For both SiFive cores, we see four groups of timing types which seem
to show to observable different latencies for the write operation,
while five groups show observable differences for the read operation.
We assume the observed latency overheads appearing for the write
operation come from the directory-based cache coherence protocol which ensures
that other existing cache block copies are invalidated on remote caches.
Regarding the T-Head core, four groups of timing types can be observed,
both for the write and read operations.
Nevertheless, we see differences between the read and
write operations due to the use of a snooping cache coherence
protocol.
In particular, a MESI protocol is used between L1-D caches with a
bypassing mechanism and a MOESI protocol is used to ensure coherence between
L1-D caches and the L2 cache.
Our interpretation is that overheads appear either due to the cache coherence, to the
snoop buffer, to store (or write) buffers, or to the load-store unit
present between cores and their L1 cache.
However, timing type latencies observed for the flush operation
are more heterogeneous than for the read and write operations.
In particular, latencies are larger when cache block has to be written
back into main memory (cache block is dirty), and latencies decrease
when cache block is located in lower levels of the cache hierarchy.
The minimum latency observed for this operation is when cache block is already
written back into main memory.

From the previous observations, we learn that the read operation
is the main leakage source on both SiFive cores,
while the dominant leakage source on the T-Head is the flush operation.
We also learn that the C910 core exposes more potentially vulnerable
behavior than the SiFive cores: first due to the differences observed between all
memory operations and, second, due to the availability of the \texttt{flush}
instruction.


\subsection{L1 Data Cache Vulnerabilities}
\label{sec:disc:l1_eval}

\subsubsection{Measurements.}
\begin{sidewaysfigure}
    \centering
    \includegraphics[width=\textwidth,keepaspectratio]{
        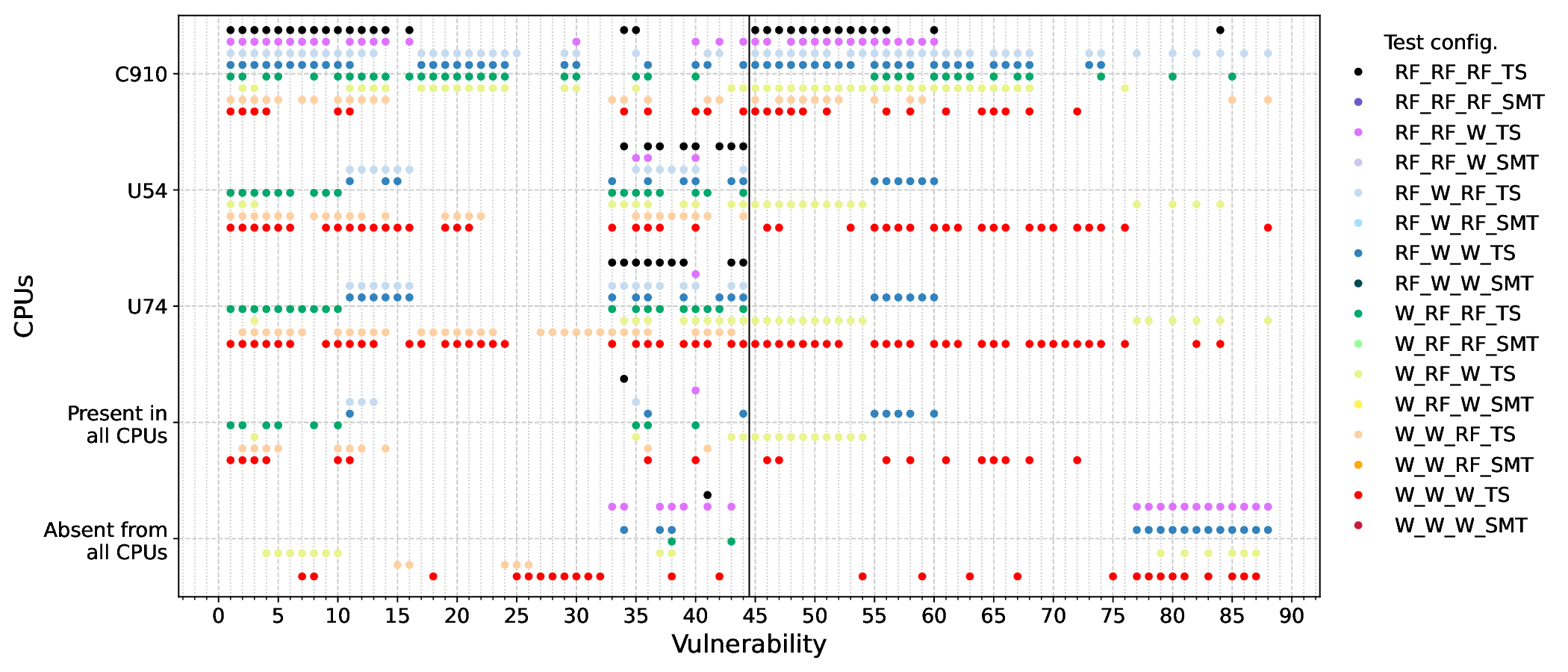}
    \caption{Evaluated timing types per vulnerability on target RISC-V SoCs:
        vulnerabilities on the left part have a read or write access as a third
        step, while vulnerabilities on the right have an invalidation as a
        third step.
        To read this plot, choose a target on the vertical axis.
        Then, for a vulnerability located on the horizontal axis, determine
        which test configuration allows to detect the vulnerability for the
        target.
        Example: vulnerability $\#1$ uses an invalidation as a first
        step.
        From all targets, markers from the top half dedicated to this
        vulnerability are only present for the C910 core.
        The top half corresponds to test configurations starting with
        \texttt{RF_}; they are test configurations where the first step
        is performed either with a \texttt{read} operation (memory access)
        or with a \texttt{flush} operation (invalidation).
        As a result, we deduce that since the flush operation is only
        accessible from U-mode on the C910, vulnerability $\#1$ will only be
        marked on the C910 for this set of test configurations.
        We can also note, for example,
        that the \texttt{W_RF_RF_TS} test configuration,
        allows to observe vulnerability $\#1$ on all targets.
    }
    \label{fig:benchmark}
\end{sidewaysfigure}

The evaluation of the three targets for the 88 strong vulnerabilities
identified by \citeA{Deng2020a} is provided by \Cref{fig:benchmark}.
On \Cref{fig:benchmark}, the presence of a vulnerability for a specific test
configuration, as described in \Cref{table:test_config}, for a given target
is indicated by a marker.
In addition, two categories indicate whether the vulnerability is present in
all evaluated CPUs or absent from all of them.

\Glspl{ctvs}~\cite{Deng2020a} are used to quantify how a target
is vulnerable by providing the ratio of vulnerabilities which are
observable on a target.
These \glspl{ctvs} are given in \Cref{table:ctvs}.
In addition, the ratio of vulnerabilities successfully observed on a
target (i.e., a success ratio),
only taking into account all valid test configurations for the latter,
is provided too in \Cref{table:ctv_op} and \Cref{table:ctv_loc}.
Valid test configurations depend on the support of flush instruction
in user mode or on the support of \gls{smt},
which leads to different maximum number of cases per target.
Thus, for each table, the success ratio for test configurations shared between
all targets or exclusive to some targets (e.g., depending on the \texttt{flush}
instruction support) is given.
This success ratio is obtained by dividing the number of successfully observed
vulnerabilities for the target with the total number of cases for the
considered category.

\begin{table}[t]
\setlength\extrarowheight{4pt}
    \caption{\small \Acrlong{ctvs}.}
    \label{table:ctvs}
    \begin{center}
        \begin{tabular}[c]{|c|c|c|c|c|c|}
            \hline
            & \textbf{C910} & \textbf{U54} & \textbf{U74}
            & \textbf{All} & \textbf{None} \\
            \hline
            \textbf{CTVS} & 70/88 & 64/88 & 77/88 & 58/88 & 6/88 \\
            \hline
        \end{tabular}
    \end{center}
\end{table}

\begin{table}[bt]
\caption{\small Ratio of vulnerabilities successfully observed for
    different running configurations: whether an attacker is present or
    not and whether the victim and the attacker run on the same core
    (time-slicing or hyper-threading) or not. Note that none of the targets
    support \gls{smt}.}
\label{table:ctv_op}
\begin{minipage}{0.50\textwidth}
\begin{subtable}[t]{\textwidth}
\centering
{
\resizebox{1\textwidth}{!}{%
\scriptsize
{\fontsize{7}{9}\selectfont
    \begin{tabular}{|c|c|c|c|c|}
    \hline
    \multirow{2}{*}{\shortstack{Core}}
    & \multicolumn{2}{c|}{\shortstack{Victim, attacker\\Same core}}
    & \multirow{2}{*}{\shortstack{Victim, attacker\\Different cores}}
    & \multirow{2}{*}{\shortstack{Victim only}}  \\ \cline{2-3}
    ~ & Time-slicing & \gls{smt} & ~ & ~ \\ \hline\hline
    C910 & 36/157 & 0 & 17/38 & 34/93 \\ \hline
    U54  & 70/157  & 0 & 24/38 & 54/93 \\ \hline
    U74  & 89/157  & 0 & 26/38 & 62/93 \\ \hline\hline
    All  & 23/157  & 0 & 14/38 & 28/93 \\ \hline
    None & 52/157  & 0 & 8/38 & 24/93 \\ \hline
    \end{tabular}
    }
    } 
} 
\caption{\small 
Shared test configurations.}
\end{subtable}
\end{minipage}
\begin{minipage}{0.50\textwidth}
\begin{subtable}[t]{\textwidth}
\centering
{
\resizebox{1\textwidth}{!}{%
\scriptsize
{\fontsize{7}{9}\selectfont
    \begin{tabular}{|c|c|c|c|c|}
    \hline
    \multirow{2}{*}{\shortstack{Core}}
    & \multicolumn{2}{c|}{\shortstack{Victim, attacker\\Same core}}
    & \multirow{2}{*}{\shortstack{Victim, attacker\\Different cores}}
    & \multirow{2}{*}{\shortstack{Victim only}}  \\ \cline{2-3}
    ~ & Time-slicing & \gls{smt} & ~ & ~ \\ \hline\hline
    C910 & 122/233 & 0 & 31/52 & 65/131 \\ \hline
    \end{tabular}
    }
    } 
} 
\caption{\small 
Exclusive test configurations.}
\end{subtable}
\end{minipage}
\end{table}

\begin{table}[bt]
\caption{\small Ratio of vulnerabilities successfully observed for
    different test configurations and memory operations. \emph{Inv.} means invalidation.}
\label{table:ctv_loc}
\begin{minipage}{0.50\textwidth}
\begin{subtable}[t]{\textwidth}
\centering
{
\resizebox{1\textwidth}{!}{%
\scriptsize
{\fontsize{7}{9}\selectfont
    \begin{tabular}{|C{2.15cm}|C{0.97cm}|C{0.97cm}|C{1.22cm}|C{1.05cm}| }
    \hline

    Core & Local Read & Local Write
    & Remote Write Inv. & Flush Inv. \\ \hline\hline

    C910 & 34/96 & 24/96 & 29/96 & 0 \\ \hline
    U54 & 62/96 & 45/96 & 41/96 & 0 \\ \hline
    U74 & 75/96 & 52/96 & 50/96 & 0 \\ \hline\hline
    All & 24/96 & 16/96 & 25/96 & 0 \\ \hline
    None & 8/96 & 32/96 & 44/96 & 0 \\ \hline
    \end{tabular}
    }
    } 
} 
\caption{\small 
Shared test configurations.}
\end{subtable}
\end{minipage}
\begin{minipage}{0.50\textwidth}
\begin{subtable}[t]{\textwidth}
\centering
{
\resizebox{1\textwidth}{!}{%
\scriptsize
{\fontsize{7}{9}\selectfont
    \begin{tabular}{|C{2.15cm}|C{0.97cm}|C{0.97cm}|C{1.22cm}|C{1.05cm}| }
    \hline

    Core & Local Read & Local Write
    & Remote Write Inv. & Flush Inv. \\ \hline\hline

    C910 & 51/80 & 49/80 & 48/80 & 70/176 \\ \hline
    \end{tabular}
    }
    } 
} 
\caption{\small 
Exclusive test configurations.}
\end{subtable}
\end{minipage}
\end{table}

\subsubsection{Observations and Preliminary Discussion.}
A first observation that can be made from \Cref{fig:benchmark} is regarding the
vulnerabilities which are absent from all boards.
In particular, absent vulnerabilities relate to observable timing
differences by evicting or invalidating addresses out of the sensitive memory
locations, which conflict with addresses in the sensitive memory locations of
interest.
This result can be interpreted as follows:
\begin{enumerate*}[label=(\roman*)]
    \item most of the vulnerabilities relying on invalidation are on the C910
        core and SiFive cores only rely on remote \texttt{write} operations
        to invalidate cache blocks, in our implementation;
    \item no significant timing difference is observed for the \texttt{write}
        operation on SiFive cores between \texttt{REMOTE_L1_CLEAN} and
        \texttt{REMOTE_L2_CLEAN}, making vulnerabilities $\#79$,
        $\#81$, and $\#83$,
        which rely on the eviction of remote-only memory
location,
        less effective.
\end{enumerate*}

Then, from \Cref{table:ctvs}, it also appears that the U74 core
exhibits more vulnerabilities than the U54.
We could interpret this as the former being more vulnerable or less secure than the latter.
However, from the more detailed view provided by \Cref{fig:benchmark},
we learn that the presence of these vulnerabilities on
one of these cores is a result of the specific benchmark run
as the vulnerabilities are only observed for one or two test
configurations.
This is supported by the similarities between both microarchitectures, in
particular for the L1 cache,  and \Cref{fig:histograms} which does not
highlight any significant difference between observable timing types. Thus, we
assume that if some vulnerabilities identified on the U74 core are false
positives, the number of vulnerabilities absent from all evaluated targets
should increase.
Depending on the previous assumption, the U74 would be either more
or less (in case previous assumption is correct) vulnerable than the C910,
but further research is needed to assess this.

Even though the C910 might present more vulnerabilities than both SiFive cores,
it appears from \Cref{table:ctv_op} and \Cref{table:ctv_loc} that, on shared
test configurations, less configurations led to observable timing differences
on the C910.
It seems that, if the C910 presents more vulnerabilities, both SiFive cores
present more ways to exploit each of their vulnerabilities.
Finally, it also clearly appears that allowing the \texttt{flush} instruction
provides ways to exploit these vulnerabilities which are not present on SiFive
cores, from user mode.

\subsection{Discussion}

From our results in~\Cref{sec:disc:timing} and~\Cref{fig:benchmark} we learn that the T-Head C910, featuring a U-mode
\texttt{flush} instruction, has more observable latency differences than
the other cores.
These differences provide additional attack surface and potentially
more exploitable vulnerabilities.
An effective way to improve the security the C910 would be to restrict access
to \texttt{flush} to the M-mode (cf.~\cite{Gerlach2023}).

Following our interpretation of cycle latencies and \gls{ctvs},
both SiFive cores, the U54 and U74, expose the same set of vulnerabilities which
is smaller than that of the T-Head core.
The absence of the \texttt{flush} instruction in U-mode definitively plays
an important role for test configurations relying on a local invalidation step.
An attacker who gains control over the kernel or firmware may still be
able to invoke \texttt{flush} in M-mode. However, these attacks are
beyond the scope of our threat model and we therefore did not study timing
of \texttt{flush} instructions on these cores.
Yet, on shared test configurations
(cf.~\Cref{table:ctv_op} and \Cref{table:ctv_loc}) both SiFive cores score substantially higher numbers
vulnerabilities than the T-Head core.
This may be due to the difference of cache coherence protocols
or other microarchitectural design differences.

Both parts of the benchmark suite, the evaluation of timing types and the
evaluation of L1-D cache vulnerabilities, provide valuable information
about timing types an adversary may exploit a RISC-V processor.  The
histograms should help the designer to identify leakage sources in their
design and to make choices to improve its resilience against cache-based
timing vulnerabilities.  Then, identifying the test configurations for
which an implementation present a given vulnerability help to assess the
security of their design and to determine which implemented countermeasures
and design choices really have an impact on their design security, e.g.,
restricting the access to \texttt{flush} instructions.

\paragraph{Limitations.}
Our benchmark does not consider virtual addresses and S-mode executions in
general. Thus, our results are representative for embedded systems
but are incomplete regarding scenarios that involve an operating system.
Depending on how virtual addresses are managed, finding conflicting addresses
in a cache set may be more difficult in these scenarios.
Moreover, we did not study systems with an L3 cache yet, which may,
depending on the availability of \texttt{flush}, present even more
timing differences than the cores studied.  Our benchmark is also dependent on 
knowledge of the target cache specifications (cf.~\Cref{tab:sbc-cache}). Evaluating implementations with missing
specifications may lead to poor results.  Finally, we only use the eviction
algorithm from~\citeA{Gruss2016b}, which may be limited depending on 
cache replacement policies or the supported access pattern detection
features.

\section{Related Work}\label{sec:rw}

\citeA{Kelsey1998} first introduced the idea of using the cache hit ratio to
perform attacks on \glspl{sbox} from symmetric encryption algorithms.
Based on this idea, \citeA{Page2002} implements a first example of cache
attack on \acrshort{des}.
\citeA{Bernstein2005} reported the extraction of an AES key from a remote
computer by observing larger timings on \glspl{sbox} due to cache block
eviction, and \citeA{Percival2005} describes how memory between threads could be
used both as a covert channel and as a side channel to attack RSA.

New ways to leverage cache timing differences are still being
proposed.
\citeA{Gruss2015} introduce cache template attacks which consist on a phase
dedicated to the cache profiling with respect to an event of interest,
followed by an exploitation phase which infers event occurrences by monitoring
the cache.
\citeA{Yan2019} illustrates that cache coherence protocols, more specifically
        directory-based protocols,
can be considered an exploitable source of timing leakage.
\citeA{Briongos2020} introduces Reload$+$Refresh which exploits the cache
replacement policies and determines if a cache block is accessed by the victim.

For eviction-based attacks, different approaches to
perform eviction or to build eviction sets exist, including
eviction-based remote Rowhammer~\cite{Kim2014} and a parameterizable eviction
strategy by \citeA{Gruss2016b}.
The latter strategy is the one used in our work.
Similarly, \citeA{Liu2015}, \citeA{Song2019} and \citeA{Vila2019} present
formal definitions and methods to build minimal eviction sets.
These approaches could be incorporated in our work by relying on conflict sets
to perform the eviction, as an alternative to our method
(see~\Cref{sec:porting}) and as originally performed by \citeA{Deng2020a}.

More recently, Lipp et al.~\cite{Lipp2021,Lipp2022} and \citeA{Kogler2023}
propose a software-based power side-channel attacks which measure power
consumption due to operations on the cache or to the bit flips occurring on
cache replacement to extract keys or to break address space randomization.
If similar software interfaces allowing to monitor the power consumption would
be available on RISC-V processors, timing measurement could be replaced in our
benchmark with power consumption monitoring operations as an attempt to detect
cache misses, similarly to what \citeA{Bertoni2005} proposed.

\citeA{Lyu2017}, \citeA{Su2021a} and \citeA{Shen2021} survey cache-based
side channel attacks and countermeasures.  \citeA{Purnal2021} analyze the
effectiveness of randomization to protect caches and \citeA{Deng2021a}
analyze the effectiveness of secure caches on ARM processors.  Our
benchmark can be used to validate hardened RISC-V designs that incorporate
these countermeasures.

\citeA{Le2023} study the security of the RISC-V architecture and
\citeA{Gerlach2023} propose benchmarks and cases to evaluate and
demonstrate vulnerabilities in selected RISC-V boards, including some of
the 88 vulnerabilities from \citeA{Deng2020a}.  \citeA{Thomas2024} provide
a fuzzing tool to detect architectural vulnerabilities in RISC-V
implementations.  Work in this direction is complementary to our approach,
providing additional points of evidence for the presence or absence of
vulnerabilities and building confidence in existing designs.

Finally, \citeA{oleksenko2023hide} propose a
model-based black-box approach to detect speculative leakage sources in
CPUs by observing traces.  \citeA{fabian2022automatic} propose a framework,
supporting formal models for composite speculation mechanisms, which is
used inside a program analysis tool to detect speculative vulnerabilities
in code snippets.  \citeA{barthe2024testing} present a testing framework to
assess given program security based on a description language dedicated to
leakage models of microarchitectural optimizations.  In comparison, our
work is a model-based white-box approach focusing on the identification of
vulnerabilities, which could inform leakage models of formal approaches to
program security.  Since we use a white-box approach, implementation
specifications help to obtain accurate results by adequately adapting the
benchmark parameters to the target. However, our work is predominantly
useful for processor developers, enabling early evaluation of a processor
implementation for cache-based timing leakage.

\section{Conclusion}\label{sec:conclusion}

In this paper we reported on porting a comprehensive benchmark suite for
cache-based timing vulnerabilities to RISC-V. We evaluate the benchmark on
three commercially available RISC-V cores and present our findings, showing
diverging leakage profiles across the processors with $65.9\%$ of
vulnerabilities present across all three processors and $6.8\%$ of
vulnerabilities being absent from all cores. In addition, $37.5\%$ of the
vulnerabilities are present for at least one shared test configuration across
all three processors. Our benchmark and evaluation artifacts are available
under an open-source license.

We anticipate that the ported benchmark will be useful for researchers as
well as commercial developers of RISC-V implementations to evaluate leakage
patterns of processors, to inform developers of potentially diverging
security risks across different implementations, and to guide the
development of strong testing and verification tools for RISC-V processors
as well as for software compiled for RISC-V. As our benchmark suite helps
to identify the different timing types and leakage patterns, our work can
also support the development of concrete and formal leakage models. We have
taken care and refactored the benchmark specifically to allow for easy
re-configuration towards microarchitectures with different cache
hierarchies, different cache set associativies, and different cache
eviction strategies, which should greatly improve reusability of our
artifacts.

In future work, we will evaluate the security of processors
from other vendors and the exploitability of our findings for S-mode and U-mode.
Regarding exploitability, we will provide proofs of concept based on the benchmark for specific
applications with different risk profiles. Furthermore, we will evaluate 
countermeasures against cache timing
vulnerabilities, e.g., cache partitioning and secure caches. We will also
investigate the impact of different
cache-coherence models on the security of open-source RISC-V cores with our benchmark. The
security of implementations of Trusted Execution Environments against
cache-based timing vulnerabilities could be another target of future work.  Finally,
we seek to extend the benchmark towards similar vulnerabilities, e.g.,
translation-lookaside buffer or transient execution vulnerabilities.

\glsaddall

\glsresetall

\ifdefined\iscameraready
\subsubsection{Acknowledgements.} This research is supported by
the CyberExcellence program of the Wallon region of Belgium under
GA \#2110186.
\fi



{\small

\printbibliography 

@Misc{BVAhead,
  author           = {{BeagleBoard.org Foundation}},
  title            = {{BeagleV-Ahead}},
  language         = {English},
  note             = {Last visit: 2025-02-21},
  url              = {https://www.beagleboard.org/boards/beaglev-ahead},
}

@Misc{BVFire,
  author           = {{BeagleBoard.org Foundation}},
  title            = {{BeagleV-Fire}},
  language         = {English},
  note             = {Last visit: 2025-02-21},
  url              = {https://www.beagleboard.org/boards/beaglev-fire},
}

@Misc{Unmatched,
  author           = {{SiFive}},
  title            = {{HiFive Unmatched}},
  language         = {English},
  note             = {Last visit: 2025-02-21},
  url              = {https://www.sifive.com/boards/hifive-unmatched},
}

@InProceedings{Lipp2018,
  author           = {Moritz Lipp and Michael Schwarz and Daniel Gruss and Thomas Prescher and Werner Haas and Anders Fogh and Jann Horn and Stefan Mangard and Paul Kocher and Daniel Genkin and Yuval Yarom and Mike Hamburg},
  booktitle        = {27th {USENIX} Security Symposium ({USENIX} Security 18)},
  title            = {Meltdown: Reading Kernel Memory from User Space},
  modificationdate = {2024-06-11T13:49:06},
  readstatus       = {skimmed},
  year             = {2018},
}

@Article{Lyu2017,
  author           = {Yangdi Lyu and Prabhat Mishra},
  journal          = {Journal of Hardware and Systems Security},
  title            = {A Survey of Side-Channel Attacks on Caches and Countermeasures},
  year             = {2017},
  number           = {1},
  pages            = {33--50},
  volume           = {2},
  doi              = {10.1007/s41635-017-0025-y},
  modificationdate = {2025-01-22T17:50:54},
  publisher        = {Springer Science and Business Media {LLC}},
}

@InProceedings{Liu2015,
  author           = {Liu, Fangfei and Yarom, Yuval and Ge, Qian and Heiser, Gernot and Lee, Ruby B.},
  booktitle        = {2015 IEEE Symposium on Security and Privacy},
  title            = {Last-Level Cache Side-Channel Attacks are Practical},
  year             = {2015},
  pages            = {605-622},
  doi              = {10.1109/SP.2015.43},
  modificationdate = {2024-11-22T15:40:20},
}

@InBook{Osvik2006,
  author           = {Osvik, Dag Arne and Shamir, Adi and Tromer, Eran},
  pages            = {1--20},
  publisher        = {Springer Berlin Heidelberg},
  title            = {Cache Attacks and Countermeasures: The Case of AES},
  year             = {2006},
  booktitle        = {Lecture Notes in Computer Science},
  doi              = {10.1007/11605805_1},
  issn             = {1611-3349},
  modificationdate = {2025-01-22T17:52:22},
}

@InProceedings{Yarom2014,
  author           = {Yuval Yarom and Katrina Falkner},
  booktitle        = {23rd USENIX Security Symposium (USENIX Security 14)},
  title            = {{FLUSH+RELOAD}: A High Resolution, Low Noise, L3 Cache {Side-Channel} Attack},
  year             = {2014},
  address          = {San Diego, CA},
  month            = aug,
  pages            = {719--732},
  publisher        = {USENIX Association},
  modificationdate = {2024-11-22T15:40:20},
  url              = {https://www.usenix.org/conference/usenixsecurity14/technical-sessions/presentation/yarom},
}

@InBook{Gruss2016,
  author           = {Gruss, Daniel and Maurice, Clémentine and Wagner, Klaus and Mangard, Stefan},
  pages            = {279--299},
  publisher        = {Springer International Publishing},
  title            = {Flush+Flush: A Fast and Stealthy Cache Attack},
  year             = {2016},
  booktitle        = {Lecture Notes in Computer Science},
  doi              = {10.1007/978-3-319-40667-1_14},
  issn             = {1611-3349},
  modificationdate = {2024-12-03T17:53:48},
}

@Book{Szefer2019,
  author           = {Szefer, Jakub},
  publisher        = {Springer International Publishing},
  title            = {Principles of Secure Processor Architecture Design},
  year             = {2019},
  creationdate     = {2024-03-01T18:33:31},
  doi              = {10.1007/978-3-031-01760-5},
  issn             = {1935-3243},
  journal          = {Synthesis Lectures on Computer Architecture},
  modificationdate = {2024-11-22T15:40:21},
}

@InProceedings{Gerlach2023,
  author           = {Gerlach, Lukas and Weber, Daniel and Zhang, Ruiyi and Schwarz, Michael},
  booktitle        = {2023 IEEE Symposium on Security and Privacy (SP)},
  title            = {A Security RISC: Microarchitectural Attacks on Hardware RISC-V CPUs},
  year             = {2023},
  pages            = {2321-2338},
  creationdate     = {2024-05-08T14:37:30},
  doi              = {10.1109/SP46215.2023.10179399},
  modificationdate = {2024-11-22T15:40:21},
}

@PhdThesis{Le2023,
  author           = {Le, Anh-Tien},
  title            = {Research of RISC-V Out-of-order Processor Cache-Based Side-Channel Attacks-Systematic Analysis, Security Models and Countermeasures},
  year             = {2023},
  creationdate     = {2024-06-11T15:06:26},
  modificationdate = {2024-11-22T15:40:21},
}

@InProceedings{Gruss2015,
  author           = {Daniel Gruss and Raphael Spreitzer and Stefan Mangard},
  booktitle        = {24th USENIX Security Symposium (USENIX Security 15)},
  title            = {Cache Template Attacks: Automating Attacks on Inclusive {Last-Level} Caches},
  year             = {2015},
  address          = {Washington, D.C.},
  month            = aug,
  pages            = {897--912},
  publisher        = {USENIX Association},
  creationdate     = {2024-06-27T15:16:39},
  modificationdate = {2024-11-22T15:40:21},
  url              = {https://www.usenix.org/conference/usenixsecurity15/technical-sessions/presentation/gruss},
}

@InProceedings{Irazoqui2016,
  author           = {Irazoqui, Gorka and Eisenbarth, Thomas and Sunar, Berk},
  booktitle        = {Proceedings of the 11th ACM on Asia Conference on Computer and Communications Security},
  title            = {Cross Processor Cache Attacks},
  year             = {2016},
  month            = may,
  publisher        = {ACM},
  series           = {ASIA CCS ’16},
  collection       = {ASIA CCS ’16},
  creationdate     = {2024-06-27T15:20:05},
  doi              = {10.1145/2897845.2897867},
  modificationdate = {2024-11-22T15:40:21},
}

@InProceedings{Deng2020a,
  author           = {Deng, Shuwen and Xiong, Wenjie and Szefer, Jakub},
  booktitle        = {Proceedings of the 25th International Conference on Architectural Support for Programming Languages and Operating Systems},
  title            = {A Benchmark Suite for Evaluating Caches’ Vulnerability to Timing Attacks},
  year             = {2020},
  month            = mar,
  publisher        = {ACM},
  series           = {ASPLOS ’20},
  collection       = {ASPLOS ’20},
  creationdate     = {2024-07-02T14:00:34},
  doi              = {10.1145/3373376.3378510},
  modificationdate = {2024-11-22T15:40:22},
}

@Article{Deng2019,
  author           = {Deng, Shuwen and Xiong, Wenjie and Szefer, Jakub},
  journal          = {Journal of Hardware and Systems Security},
  title            = {Analysis of Secure Caches Using a Three-Step Model for Timing-Based Attacks},
  year             = {2019},
  issn             = {2509-3436},
  month            = nov,
  number           = {4},
  pages            = {397--425},
  volume           = {3},
  creationdate     = {2024-07-02T15:41:37},
  doi              = {10.1007/s41635-019-00075-9},
  modificationdate = {2024-11-22T15:40:22},
  publisher        = {Springer Science and Business Media LLC},
}

@Book{Harris2022,
  author           = {Harris, Sarah L.},
  editor           = {David Money Harris},
  publisher        = {Morgan Kaufmann},
  title            = {Digital design and computer architecture: {RISC-V} edition},
  year             = {2022},
  address          = {Cambridge},
  note             = {Includes index},
  creationdate     = {2024-07-08T14:03:28},
  modificationdate = {2024-11-28T16:03:00},
  pagetotal        = {1564},
  ppn_gvk          = {1796239658},
}

@Misc{Deng2021a,
  author           = {Deng, Shuwen and Matyunin, Nikolay and Xiong, Wenjie and Katzenbeisser, Stefan and Szefer, Jakub},
  title            = {Evaluation of Cache Attacks on Arm Processors and Secure Caches},
  year             = {2021},
  copyright        = {arXiv.org perpetual, non-exclusive license},
  creationdate     = {2024-10-04T14:34:36},
  doi              = {10.48550/ARXIV.2106.14054},
  modificationdate = {2024-11-22T15:40:22},
  publisher        = {arXiv},
}

@InProceedings{Yan2019,
  author           = {Yan, Mengjia and Sprabery, Read and Gopireddy, Bhargava and Fletcher, Christopher and Campbell, Roy and Torrellas, Josep},
  booktitle        = {2019 IEEE Symposium on Security and Privacy (SP)},
  title            = {Attack Directories, Not Caches: Side Channel Attacks in a Non-Inclusive World},
  year             = {2019},
  month            = may,
  publisher        = {IEEE},
  creationdate     = {2024-10-22T16:05:39},
  doi              = {10.1109/sp.2019.00004},
  modificationdate = {2024-11-22T15:40:22},
}

@Article{Su2021a,
  author           = {Su, Chao and Zeng, Qingkai},
  journal          = {Security and Communication Networks},
  title            = {Survey of CPU Cache-Based Side-Channel Attacks: Systematic Analysis, Security Models, and Countermeasures},
  year             = {2021},
  issn             = {1939-0114},
  month            = jun,
  pages            = {1--15},
  volume           = {2021},
  creationdate     = {2024-10-23T16:47:36},
  doi              = {10.1155/2021/5559552},
  editor           = {Nicopolitidis, Petros},
  modificationdate = {2024-11-22T15:40:22},
  publisher        = {Hindawi Limited},
}

@InProceedings{Purnal2021,
  author           = {Purnal, Antoon and Giner, Lukas and Gruss, Daniel and Verbauwhede, Ingrid},
  booktitle        = {2021 IEEE Symposium on Security and Privacy (SP)},
  title            = {Systematic Analysis of Randomization-based Protected Cache Architectures},
  year             = {2021},
  month            = may,
  pages            = {987--1002},
  publisher        = {IEEE},
  creationdate     = {2024-10-25T14:08:54},
  doi              = {10.1109/sp40001.2021.00011},
  modificationdate = {2024-11-22T15:40:22},
}

@InProceedings{Bernstein2005,
  author           = {Daniel J. Bernstein},
  title            = {Cache-timing attacks on AES},
  year             = {2005},
  creationdate     = {2024-10-25T14:16:40},
  modificationdate = {2024-11-22T15:40:22},
  url              = {https://cr.yp.to/antiforgery/cachetiming-20050414.pdf},
}

@Article{Percival2005,
  author           = {Percival, Colin},
  title            = {Cache missing for fun and profit},
  year             = {2005},
  month            = aug,
  creationdate     = {2024-10-25T15:10:56},
  modificationdate = {2025-01-13T17:34:02},
  url              = {https://www.daemonology.net/hyperthreading-considered-harmful/},
}

@InProceedings{Lipp2016,
  author           = {Moritz Lipp and Daniel Gruss and Raphael Spreitzer and Cl{\'e}mentine Maurice and Stefan Mangard},
  booktitle        = {25th USENIX Security Symposium},
  title            = {{ARMageddon}: Cache Attacks on Mobile Devices},
  year             = {2016},
  month            = aug,
  pages            = {549--564},
  publisher        = {USENIX Association},
  creationdate     = {2024-11-04T16:13:11},
  modificationdate = {2024-11-22T15:40:22},
  url              = {https://www.usenix.org/conference/usenixsecurity16/technical-sessions/presentation/lipp},
}

@InProceedings{Bertoni2005,
  author           = {Bertoni, G. and Zaccaria, V. and Breveglieri, L. and Monchiero, M. and Palermo, G.},
  booktitle        = {International Conference on Information Technology: Coding and Computing (ITCC’05) - Volume II},
  title            = {AES power attack based on induced cache miss and countermeasure},
  year             = {2005},
  pages            = {586--591 Vol. 1},
  publisher        = {IEEE},
  creationdate     = {2024-11-05T17:09:23},
  doi              = {10.1109/itcc.2005.62},
  modificationdate = {2024-11-22T15:40:22},
}

@PhdThesis{Purnal2023a,
  author           = {Purnal, Antoon and Verbauwhede, Ingrid},
  title            = {Cache Side-Channel Attacks on Existing and Emerging Computing Platforms},
  language         = {eng},
  creationdate     = {2024-11-05T17:43:46},
  modificationdate = {2024-11-05T17:46:51},
  year             = {2023},
}

@InProceedings{Vila2019,
  author           = {Vila, Pepe and Kopf, Boris and Morales, Jose F.},
  booktitle        = {2019 IEEE Symposium on Security and Privacy (SP)},
  title            = {Theory and Practice of Finding Eviction Sets},
  year             = {2019},
  month            = may,
  pages            = {39--54},
  publisher        = {IEEE},
  comment-hgwk     = {https://github.com/cgvwzq/evsets},
  creationdate     = {2024-11-05T17:58:38},
  doi              = {10.1109/sp.2019.00042},
  modificationdate = {2024-11-22T15:40:22},
}

@InProceedings{Song2019,
  author           = {Wei Song and Peng Liu},
  booktitle        = {22nd International Symposium on Research in Attacks, Intrusions and Defenses (RAID 2019)},
  title            = {Dynamically Finding Minimal Eviction Sets Can Be Quicker Than You Think for {Side-Channel} Attacks against the {LLC}},
  year             = {2019},
  address          = {Chaoyang District, Beijing},
  month            = sep,
  pages            = {427--442},
  publisher        = {USENIX Association},
  creationdate     = {2024-11-06T14:14:55},
  modificationdate = {2024-11-22T15:40:22},
  url              = {https://www.usenix.org/conference/raid2019/presentation/song},
}

@InProceedings{Shen2021,
  author           = {Shen, Chaoqun and Chen, Congcong and Zhang, Jiliang},
  booktitle        = {Proceedings of the 26th Asia and South Pacific Design Automation Conference},
  title            = {Micro-architectural Cache Side-Channel Attacks and Countermeasures},
  year             = {2021},
  month            = jan,
  pages            = {441--448},
  publisher        = {ACM},
  series           = {ASPDAC ’21},
  collection       = {ASPDAC ’21},
  creationdate     = {2024-11-07T16:49:06},
  doi              = {10.1145/3394885.3431638},
  modificationdate = {2024-11-22T15:40:22},
}

@Article{Thomas2024,
  author           = {Thomas, Fabian and Hetterich, Lorenz and Zhang, Ruiyi and Weber, Daniel and Gerlach, Lukas and Schwarz, Michael},
  title            = {{RISCVuzz}: Discovering Architectural {CPU} Vulnerabilities via Differential Hardware Fuzzing},
  year             = {2024},
  creationdate     = {2024-12-03T16:10:40},
  howpublished     = {\url{https://ghostwriteattack.com/}},
  modificationdate = {2024-12-03T16:39:05},
}

@InProceedings{Gruss2016b,
  author           = {Gruss, Daniel and Maurice, Clémentine and Mangard, Stefan},
  title            = {Rowhammer.js: A Remote Software-Induced Fault Attack in JavaScript},
  year             = {2016},
  month            = {07},
  pages            = {300-321},
  creationdate     = {2024-12-03T17:59:13},
  doi              = {10.1007/978-3-319-40667-1_15},
  modificationdate = {2024-12-03T17:59:13},
}

@InProceedings{Kogler2023,
  author           = {Kogler, Andreas and Juffinger, Jonas and Giner, Lukas and Gerlach, Lukas and Schwarzl, Martin and Schwarz, Michael and Gruss, Daniel and Mangard, Stefan},
  booktitle        = {USENIX Security},
  title            = {{Collide+Power: Leaking Inaccessible Data with Software-based Power Side Channels}},
  year             = {2023},
  creationdate     = {2025-01-12T21:19:45},
  modificationdate = {2025-01-12T21:19:45},
}

@Misc{Page2002,
  author           = {D. Page},
  howpublished     = {Cryptology {ePrint} Archive, Paper 2002/169},
  title            = {Theoretical Use of Cache Memory as a Cryptanalytic Side-Channel},
  year             = {2002},
  creationdate     = {2025-01-13T15:39:11},
  modificationdate = {2025-01-13T15:39:11},
  url              = {https://eprint.iacr.org/2002/169},
}

@InBook{Kelsey1998,
  author           = {Kelsey, John and Schneier, Bruce and Wagner, David and Hall, Chris},
  pages            = {97--110},
  publisher        = {Springer Berlin Heidelberg},
  title            = {Side channel cryptanalysis of product ciphers},
  year             = {1998},
  booktitle        = {Computer Security — ESORICS 98},
  creationdate     = {2025-01-13T15:47:42},
  doi              = {10.1007/bfb0055858},
  issn             = {1611-3349},
  modificationdate = {2025-01-13T15:47:42},
}

@InProceedings{Briongos2020,
  author           = {Samira Briongos and Pedro Malagon and Jose M. Moya and Thomas Eisenbarth},
  booktitle        = {29th USENIX Security Symposium (USENIX Security 20)},
  title            = {{RELOAD+REFRESH}: Abusing Cache Replacement Policies to Perform Stealthy Cache Attacks},
  year             = {2020},
  month            = aug,
  pages            = {1967--1984},
  publisher        = {USENIX Association},
  creationdate     = {2025-01-14T11:22:30},
  modificationdate = {2025-01-14T11:22:30},
  url              = {https://www.usenix.org/conference/usenixsecurity20/presentation/briongos},
}

@InProceedings{Ren2021,
  author           = {Ren, Xida and Moody, Logan and Taram, Mohammadkazem and Jordan, Matthew and Tullsen, Dean M. and Venkat, Ashish},
  booktitle        = {2021 ACM/IEEE 48th Annual International Symposium on Computer Architecture (ISCA)},
  title            = {I See Dead µops: Leaking Secrets via Intel/AMD Micro-Op Caches},
  year             = {2021},
  pages            = {361-374},
  creationdate     = {2025-01-22T17:42:54},
  doi              = {10.1109/ISCA52012.2021.00036},
  modificationdate = {2025-01-22T17:45:27},
}

@InProceedings{Lipp2020,
  author           = {Lipp, Moritz and Had\v{z}i\'{c}, Vedad and Schwarz, Michael and Perais, Arthur and Maurice, Cl\'{e}mentine and Gruss, Daniel},
  booktitle        = {Proceedings of the 15th ACM Asia Conference on Computer and Communications Security},
  title            = {Take A Way: Exploring the Security Implications of AMD's Cache Way Predictors},
  year             = {2020},
  address          = {New York, NY, USA},
  pages            = {813–825},
  publisher        = {Association for Computing Machinery},
  series           = {ASIA CCS '20},
  creationdate     = {2025-01-22T17:48:15},
  doi              = {10.1145/3320269.3384746},
  modificationdate = {2025-01-22T17:49:40},
  numpages         = {13},
}

@InProceedings{Rauscher2025,
  author           = {Fabian Rauscher and Carina Fiedler and Andreas Kogler and Daniel Gruss},
  booktitle        = {Network and Distributed System Security Symposium (NDSS) 2025},
  title            = {A Systematic Evaluation of Novel and Existing Cache Side Channels},
  year             = {2025},
  month            = feb,
  note             = {Network and Distributed System Security Symposium 2025 :
      NDSS 2025},
  creationdate     = {2025-01-23T16:48:58},
  day              = {23},
  doi              = {10.14722/ndss.2025.23253},
  language         = {English},
  modificationdate = {2025-01-23T16:48:58},
  url              = {https://www.ndss-symposium.org/ndss2025/},
}

@InProceedings{Kim2014,
  author           = {Kim, Yoongu and Daly, Ross and Kim, Jeremie and Fallin, Chris and Lee, Ji Hye and Lee, Donghyuk and Wilkerson, Chris and Lai, Konrad and Mutlu, Onur},
  booktitle        = {2014 ACM/IEEE 41st International Symposium on Computer Architecture (ISCA)},
  title            = {Flipping bits in memory without accessing them: An experimental study of DRAM disturbance errors},
  year             = {2014},
  pages            = {361-372},
  comment-hgwk     = {Rowhammer paper},
  creationdate     = {2025-02-09T20:15:57},
  doi              = {10.1109/ISCA.2014.6853210},
  modificationdate = {2025-02-09T20:16:22},
}

@InProceedings{Lipp2022,
  author           = {Moritz Lipp and Daniel Gruss and Michael Schwarz},
  booktitle        = {31st USENIX Security Symposium (USENIX Security 22)},
  title            = {{AMD} Prefetch Attacks through Power and Time},
  year             = {2022},
  address          = {Boston, MA},
  month            = aug,
  pages            = {643--660},
  publisher        = {USENIX Association},
  comment-hgwk     = {Meltdown-Power},
  creationdate     = {2025-02-09T22:02:39},
  modificationdate = {2025-02-09T22:17:41},
  url              = {https://www.usenix.org/conference/usenixsecurity22/presentation/lipp},
}

@InProceedings{Lipp2021,
  author           = {Lipp, Moritz and Kogler, Andreas and Oswald, David and Schwarz, Michael and Easdon, Catherine and Canella, Claudio and Gruss, Daniel},
  booktitle        = {2021 IEEE Symposium on Security and Privacy (SP)},
  title            = {PLATYPUS: Software-based Power Side-Channel Attacks on x86},
  year             = {2021},
  pages            = {355-371},
  creationdate     = {2025-02-09T22:06:44},
  doi              = {10.1109/SP40001.2021.00063},
  modificationdate = {2025-02-09T22:17:53},
}

@inproceedings{barthe2024testing,
  title={Testing side-channel security of cryptographic implementations against future microarchitectures},
  author={Barthe, Gilles and B{\"o}hme, Marcel and Cauligi, Sunjay and Chuengsatiansup, Chitchanok and Genkin, Daniel and Guarnieri, Marco and Mateos Romero, David and Schwabe, Peter and Wu, David and Yarom, Yuval},
  booktitle={Proceedings of the 2024 on ACM SIGSAC Conference on Computer and Communications Security},
  pages={1076--1090},
  year={2024}
}

@inproceedings{oleksenko2023hide,
  title={Hide and Seek with Spectres: Efficient discovery of speculative information leaks with random testing},
  author={Oleksenko, Oleksii and Guarnieri, Marco and K{\"o}pf, Boris and Silberstein, Mark},
  booktitle={2023 IEEE Symposium on Security and Privacy (SP)},
  pages={1737--1752},
  year={2023},
  organization={IEEE}
}

@inproceedings{fabian2022automatic,
  title={Automatic detection of speculative execution combinations},
  author={Fabian, Xaver and Guarnieri, Marco and Patrignani, Marco},
  booktitle={Proceedings of the 2022 ACM SIGSAC Conference on Computer and Communications Security},
  pages={965--978},
  year={2022}
}
}

\end{document}